\documentclass[conference]{IEEEtran}
\IEEEoverridecommandlockouts
\usepackage{cite}
\usepackage{amsmath,amssymb,amsfonts}
\usepackage{algorithmic}
\usepackage{graphicx}
\usepackage{textcomp}
\usepackage{xcolor}

\usepackage{hyperref}
\usepackage{cleveref}

\usepackage{framed,enumitem}

\usepackage{tikz}
\usepackage{textcomp}
\usepackage{lipsum}

\newcommand\copyrighttext{%
  \footnotesize \textcopyright 2021 IEEE. Personal use of this material is permitted.
  Permission from IEEE must be obtained for all other uses, in any current or future
  media, including reprinting/republishing this material for advertising or promotional
  purposes, creating new collective works, for resale or redistribution to servers or
  lists, or reuse of any copyrighted component of this work in other works.
  DOI: \href{https://doi.org/10.1109/CBI52690.2021.10049}{10.1109/CBI52690.2021.10049}}
\newcommand\copyrightnotice{%
\begin{tikzpicture}[remember picture,overlay]
\node[anchor=south,yshift=10pt] at (current page.south) {\fbox{\parbox{\dimexpr\textwidth-\fboxsep-\fboxrule\relax}{\copyrighttext}}};
\end{tikzpicture}%
}

\usepackage[numbers]{natbib}

\def\BibTeX{{\rm B\kern-.05em{\sc i\kern-.025em b}\kern-.08em
    T\kern-.1667em\lower.7ex\hbox{E}\kern-.125emX}}
\begin{document}

\title{Towards a Reference Architecture for Future Industrial Internet of Things Networks
}

\author{\IEEEauthorblockN{1\textsuperscript{st} Dominik Martin}
\IEEEauthorblockA{\textit{Karlsruhe Institute of Technology} \\
Karlsruhe, Germany \& \\
\textit{Trelleborg Sealing Solutions Germany GmbH}\\
Stuttgart, Germany \\
dominik.martin@kit.edu}
\and
\IEEEauthorblockN{2\textsuperscript{nd} Niklas Kühl}
\IEEEauthorblockA{\textit{Karlsruhe Institute of Technology} \\
Karlsruhe, Germany \&\\
\textit{IBM}\\
Ehningen, Germany \\
niklas.kuehl@kit.edu}
\and
\IEEEauthorblockN{3\textsuperscript{rd} Marcel Schwenk}
\IEEEauthorblockA{\textit{Karlsruhe Institute of Technology} \\
Karlsruhe, Germany \\
office@ksri.kit.edu\\
 \\
 }
}

\maketitle
\copyrightnotice

\begin{abstract}
With the continuing decrease of sensor technology prices as well as the increase of communication and analytical capabilities of modern internet of things devices, the continuously generated amount of data is constantly growing. Various use cases show the untapped potential of this data for new business models. However, conventional industrial IT networks of traditional manufacturing companies can hardly meet the modern requirements emerging with today's and future industrial internet of things applications. Outdated and rigid network infrastructures are one of the main reasons for hesitant innovation efforts and cross-organizational collaborations as well as the slow adoption of modern business models by traditional manufacturing companies.
Following the design science research paradigm, our work contributes by elaborating on a comprehensive list of requirements for future industrial internet of things networks from a theoretical and practical perspective as well as a proposed reference architecture acting as a blueprint for future implementations.
\end{abstract}

\begin{IEEEkeywords}
Industrial Internet of Things, IT network, Manufacturing, Design Science Research, Reference Architecture
\end{IEEEkeywords}

\section{Introduction}

Technological and cost-related developments in the field of sensor technology are one of the drivers of the internet of things \cite{Wortmann2015} or so-called cyber-physical systems \cite{Lasi2014}. Additionally, the vast number of networking opportunities and developments in the field of communication technology are crucial for the dissemination of this sensor data. Thus, more and more industrial assets such as production machines are equipped with a wide variety of sensors that collect data about the production process, environmental conditions, characteristics of, e.g., raw materials as well as the quality of the end products. In addition, the machines themselves produce a range of production parameters, such as cycle times, or error logs. 

Research shows that industrial data bears enormous potential for the exploitation of additional value: \citet{HunkeEngel2018_1000084912}, for instance, show that innovative services based on data and analytics are a viable option to gain competitive advantages over competitors. Furthermore, \citet{Vossing2019} demonstrate how machine data in a manufacturing environment can be leveraged to improve production performance by simply providing process information and actionable information to operators. In contrast to internal optimization, machine data can also be used to create new services for external actors. \citet{Martin2021} show that especially sensor data are valuable inputs of so-called virtual sensors, which are able to capture non-measurable phenomena (e.g., the wear of machines) to support decision making (e.g., to schedule maintenance) or enable data-driven services (e.g., condition monitoring or predictive maintenance services). Thus, sensor or process data from production, if shared with other actors, can be leveraged to build a holistic information model that may enable value co-creation among various actors in a value creation ecosystem \cite{Martin2019c, MartinKuehlBischhoffshausen2020_1000118627, Bose2019}.

However, this potentially valuable data is sometimes not even used for internal purposes---and in particular it is not shared with customers or other external players. Due to concerns about sharing sensitive information and intellectual property (IP) about the own production with competitors \cite{Dietz2020, Matt2018}, but also due to the lack of flexibility of today's industrial IT infrastructure (including the network) revealing data in a targeted manner \cite{OpenNetworkingFoundation2012}, this potential remains largely untapped.


Beyond these considerations, there are also a number of unfulfilled technical requirements that prevent companies from sharing their data. These include requirements from the field of network technology, such as security and scalability. Furthermore, industrial internet of things (IIoT) applications get more and more modularized and become controlled by internal or external IT services. Customers increasingly demand customized  products or desire real time tracking through production. Industrial IT networks are thus expanding to meet new requirements such as accessibility and adaptability. To respond to these emerging requirements, the control and management functions of IT networks must be closely linked to the underlying applications and their requirements. Due to the heterogeneity and inflexibility of traditional network architectures this is only possible to a limited extent. 

To tackle this challenge, the article at hand gathers requirements in a systematic way and derives actionable principles which guide the design of a sustainable network architecture for the manufacturing environment. Thus, the following research question (RQ) and subquestions (SQs) guide our research:

\emph{RQ:} How could an IT network infrastructure for the manufacturing environment look like to support future IIoT use cases?

\emph{SQ1:} What are the requirements for the IT infrastructure of a manufacturing network in the context of future IIoT use cases?

\emph{SQ2:} Which actionable guidelines in the form of design principles address these design requirements and inform the development of a IIoT manufacturing network architecture?

\section{Related Work}
\label{sec:relatedwork}

\subsection{Software Defined Networking}

\begin{figure}[htbp]
    \centering
	\includegraphics[width=1\linewidth]{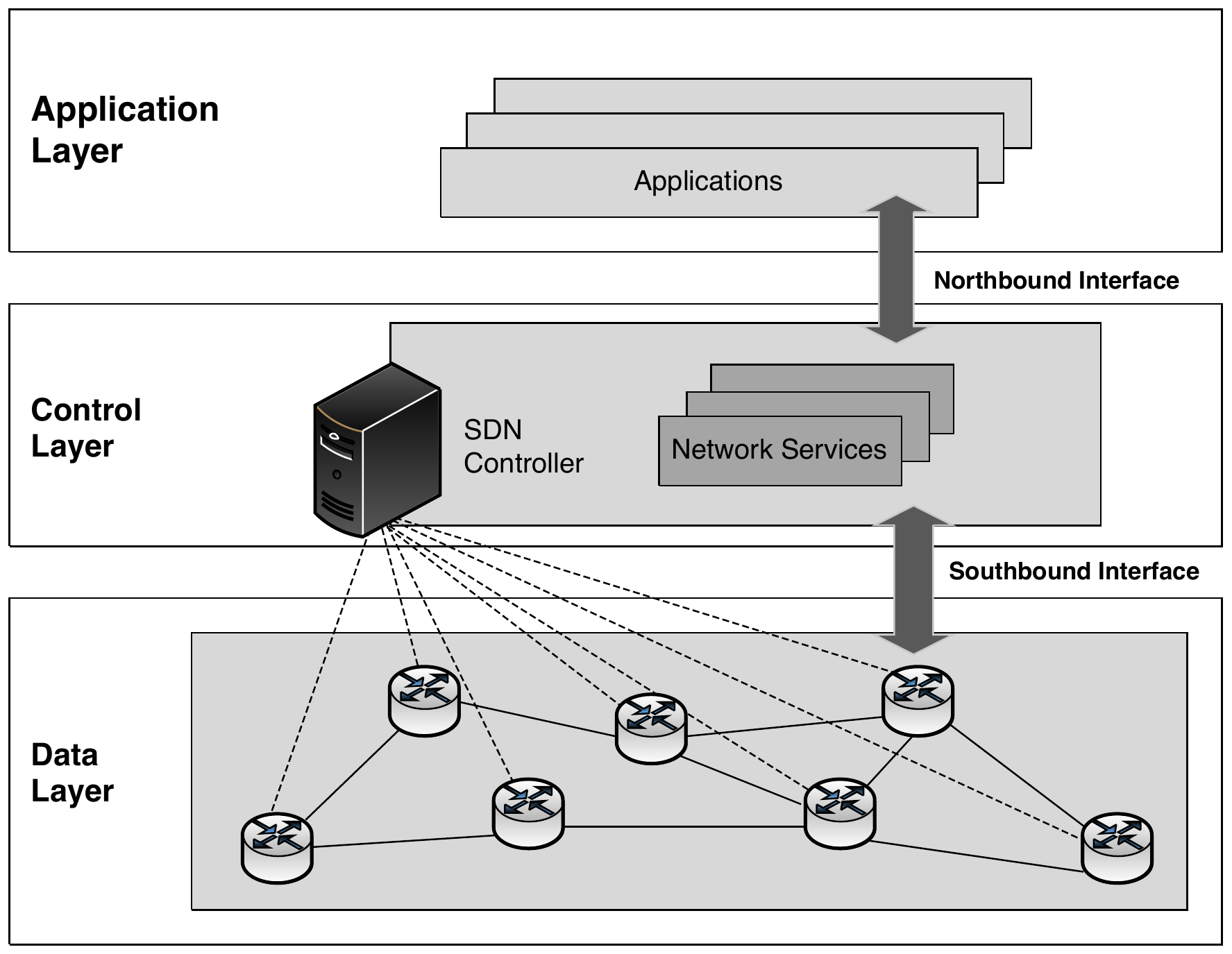}
	\caption{Software defined networking layers based on \citet{OpenNetworkFoundation2014}}
	\label{fig:software-defined networking}
\end{figure}

Software defined networking (SDN) has emerged as a new networking paradigm with the basic concept of decoupling the network control layer from the network data (i.e., infrastructure) layer. This separation, thus, enables direct programmability of network functions and reduces the effort required for network control and management.
\Cref{fig:software-defined networking} depicts the typical SDN architecture \cite{OpenNetworkFoundation2014}, which can be divided into the data layer, the control layer and the application layer.

The \emph{data layer}, also known as the forwarding layer, includes software or hardware switches and other physical devices. In general, the data layer is responsible for forwarding packets within the network based on rules defined and distributed by an SDN controller at the control layer.
The \emph{control layer} contains the logic, protocols and control functions of the network that controls forwarding at the data layer. The control layer manages network flows through switches and routers via the southbound interface. At the same time, the northbound interface is utilized for communication with the application layer. Based on the global network view of the controller, data layer devices can be controlled by providing rules for data routing.
The \emph{application layer} consists of software applications that use the northbound interface to receive an abstract network view as well as to control the behavior of the network through SDN controllers. Using this interface, applications can orchestrate the data layer to perform complex tasks such as topology detection, load balancing, enforcing firewall rules and more.

A number of articles \cite{Ahmed2018, Henneke2016, Kalman2015} deal with the general applicability of software defined networking in the context of industrial IT networks. Based on this, \citet{Ehrlich2018,Kobzan2020, Madiwalar2019} investigate to what extent SDN can meet specific requirements in the context of IIoT use cases. Especially in production environments---with long machine life cycles---it is important that old and new network technology can be used in parallel without causing interoperability issues and security threads. Therefore, \citet{Amin2018} study whether SDN can be integrated into an existing network, which problems arise and which possibilities might allow to operate legacy network areas and new areas controlled by SDN in combination. \citet{Ahmed2017} was the first presenting an architecture for industrial networks based on SDN, however, with the limitation that only the aspect of real-time communication was considered.

\begin{table*}[!htbp]
\centering
\caption{Overview of Design Science Research project choices}
\label{tab:designoverview}
\begin{tabular}{|r|l|}
\hline
\textbf{Real-world problem:}                                                                                    & \begin{tabular}[c]{@{}l@{}}Lack of reference architectures for IIoT networks\end{tabular}                                                                          \\ \hline
\textbf{Kernel theories:}                                                                                         & \begin{tabular}[c]{@{}l@{}}Software defined networking (SDN), \\ network function virtualization (NFV)\end{tabular}                                                                                                                                          \\ \hline
\textbf{\begin{tabular}[c]{@{}r@{}}Artifact type according \\ to \citet{Peffers2012}:\end{tabular}}           & Model                                                                                                               \\ \hline
\textbf{\begin{tabular}[c]{@{}r@{}}Evaluation objective according \\ to \citet{pries2008strategies}:\end{tabular}} & \begin{tabular}[c]{@{}l@{}}General \\ Feasibility\end{tabular}                                                                                                       \\ \hline
\textbf{\begin{tabular}[c]{@{}r@{}}Evaluation type according \\ to  \citet{Venable2016}:\end{tabular}}         & \begin{tabular}[c]{@{}l@{}}Episode 1: Artificial\\ Episode 2: Naturalistic\end{tabular}                                                                              \\ \hline
\textbf{\begin{tabular}[c]{@{}r@{}}Evaluation method according \\ to \citet{Peffers2012}:\end{tabular}}       & \begin{tabular}[c]{@{}l@{}}Episode 1: Logical argument\\ Episode 2: Expert evaluation\end{tabular}                                                                   \\ \hline
\textbf{\begin{tabular}[c]{@{}r@{}}Contribution according \\ to \citet{Gregor2013}:\end{tabular}}          & \begin{tabular}[c]{@{}l@{}}Improvement; application of known solution (SDN, NFV) \\ to novel problem (IIoT network architecture)\end{tabular} \\ \hline
\end{tabular}
\end{table*}

\subsection{Network Function Virtualization}
In addition to SDN, the trend towards virtualization of network functions bears huge potential to improve the provision of networked applications \cite{etsi2014network}. The idea behind network function virtualization (NFV) is the virtualization of network services, such as firewalls, which are usually deployed as dedicated hardware systems within the network \cite{Wood2015}. 

Virtualized network functions, such as data traffic analysis, the provision of firewall mechanisms and other services describe software implementations of network functions running on NFV infrastructure. This infrastructure consists of hardware and software components which can be distributed over several physical locations. A virtualization layer abstracts the hardware resources and decouples their computing, storage and network functionalities.

Compared to the common practice, NFV offers advantages in terms of flexibility, scalability and resilience of resources \cite{Wood2015}. Since network elements are no longer a collection of integrated hardware and software components, both elements can be developed independently. In addition, the separation of hardware and software ensures that they can perform different functions at different times and scale dynamically and with a finer granularity, according to the capacity required by the actual data traffic.

\citet{Wood2015} show that NFV and SDN can complement each other and, therefore,  can be used in combination. In addition, \citet{Reynaud2016} summarize different possibilities and configurations how the concepts SDN and NFV can lead to more flexible and scalable network architectures, but also to potential weaknesses for attacks.

\section{Research Design}
\label{sec:researchdesign}

Following the design science research paradigm according to \citet{1885-32762} and the guidelines proposed by \citet{Peffers2008}, we address the problem of a lacking future orientation of industrial IoT networks in the manufacturing environment. Design science research aims to guide the creation of innovative artifacts, accordingly its rigorous construction and evaluation as well as to demonstrate its utility for practical applications \cite{Peffers2012, Venable2016, 1885-32762}. The development of design knowledge which arises from the creation of artifacts is of high relevance for both research and practice \cite{Hevner2004}. Furthermore, existing theory can serve, in the form of kernel theories, as justificatory knowledge supporting the design \cite{1885-32762}. In particular, design principles derived from kernel theories may guide the implementation of an artifact \cite{walls1992building}. Furthermore, this theorizing process serves to formalize design knowledge \cite{Gregor2013}.

Thus, we identify requirements (DRs) for future IIoT networks from scientific literature based on a structured literature review according to \citet{Webster2002}. In addition, we extend the list of requirements with practical ones resulting from industrial use cases that have been identified by a panel of experts and classified as relevant for the future. Utilizing aspects derived from the kernel theories SDN and NFV we develop design principles to meet the identified design requirements. These design principles are then transformed into an artifact via actionable design features (DFs). The artifact represents a model, which according to \citet{Peffers2012} depicts a ``simplified representation of [the] reality documented using a formal notation or language''.

The artifact's evaluation is described in two evaluation episodes. First, through an artificial logical argumentation we show that the identified requirements are fulfilled. Second, by conducting a naturalistic evaluation through expert interviews in the field, we describe its applicability. \Cref{tab:designoverview} depicts a summary of the design science research choices made for a better overview. 
As a result, our work contributes by elaborating on a comprehensive list of requirements for industrial IoT networks from a theoretical and practical perspective as well as by an architecture acting as a blueprint for future implementations.

\section{Artifact Description}

As described in \Cref{sec:researchdesign}, we identify requirements for future IIoT networks in two different ways: First, we conduct a structured literature review according to \citet{Webster2002} by leveraging Google Scholar as search engine. Utilizing the search terms `future "industrial network" security requirements', `industry 4.0 it "network requirements"', `"smart factory" network management', `"industry 4.0" "use case requirements"', `future "industrial network" requirements', `"smart factory" network requirements', and `"industrial internet of things" requirements' we collect potentially relevant articles after 2010 to focus on recent ones. After analyzing the titles and abstracts, 29 out of 1400 papers are filtered, from which the requirements are extracted.

Second, in addition to the requirements from scientific literature, we collect further requirements linked to practical use cases, which are defined in an expert workshop \cite{tremblay2010focus}. A panel consisting of four experts from the areas of operations and network administration identifies three different exemplary future use cases and derives a set of requirements from them. The three selected use cases cover a visual inspection camera temporarily mapped to a specific machine, direct machine access from outside the network and an autonomous transport system that interacts with machines.

\Cref{tab:requirements} (Appendix) summarizes 44 requirements derived from literature and practice. These specific requirements are transformed into abstract design requirements in a subsequent step. The mapping of these particular requirements to their respective design requirements is also listed in \Cref{tab:requirements}. The abstract design requirements are listed below:

\begin{itemize}[leftmargin=2.65em]

\item [DR1:] \emph{Quality of Service}

Individual applications may have different requirements in regard to the service quality and performance. They may serve different purposes or use different communication patterns and, thus, may require different quality of service (QoS) parameters (e.g., latency, jitter, bandwidth, or loss rate). Traffic prioritization and resource control mechanisms need to be provided by the underlying network infrastructure. 

\item [DR2:] \emph{Non-inferring flows} 

Multiple applications may use the same network infrastructure in parallel without interfering with each other. If the same physical network infrastructure is to be used for time-critical (i.e., real-time capable) and non-time-critical communication traffic, the network service must ensure that no interference occurs.

\item [DR3:] \emph{Network segmentation} 

The network has to ensure that IT devices can be grouped into network segments which are connected by clearly defined transitions and can have individual security levels. Due to the lack of cryptographic security mechanisms in common industrial communication protocols and the increasing number of networking devices, security must be enforceable through network restrictions.

\item [DR4:] \emph{Reliability}

The network service ensures that messages are transmitted with a desired reliability. This implies, for example, that all packets are delivered complete and in the intended order. Thus, the network architecture must be designed in a way that a malfunction of individual components, such as servers or switches, does not lead to losses or even a failure of the entire network.

\item [DR5:] \emph{Confidentiality}

To ensure that unauthorized users, services or devices cannot access applications or other devices, mechanisms for user authentication must be provided. This authentication is intended to provide role-based user authorizations.

\item [DR6:] \emph{Integrity}

The network must ensure that unauthorised modification of information is prevented. This cannot be done typically due to the nature of data transmission mechanisms. Technical measures to ensure integrity therefore aim to be able to identify faulty data as such and, if necessary, to carry out a new data transmission.

\item [DR7:] \emph{Availability}

Availability describes the ratio of the time within a time period in which the system is operationally available for its actual purpose. Thus, the network must ensure that withholding of information or resources is prevented at any time.

\item [DR8:] \emph{Retrofit}

Due to long life cycles of production assets and, thus, the potentially outdated hardware, it is necessary that existing assets can be integrated into the network. The network infrastructure must provide compatibility mechanisms to retrofit ageing devices and assets.

\item [DR9:] \emph{Reconfigurability}

The network infrastructure allows new devices to be added to the existing network dynamically and without high operational effort. Setup information which may be required to add the device should be exchanged automatically via the network. In order to be able to react to changing requirements, such as changing devices or components (e.g., servers, switches, machines), it must be possible to reconfigure the network at runtime without affecting other components.

\item [DR10:] \emph{Accessibility}

In order to monitor networked assets as well as to access them in case of issues, remote access must be enabled. In addition, access to assets for streaming machine data by authorized users or user roles (inside and outside the network) should be possible.

\item [DR11:] \emph{Adaptability}

An industrial network has a very high expected lifetime due to the high setup costs and the spread of components. To be able to react to developments in the field of communication protocols, requirements, security and many more, individual components must be kept as adaptable as possible.

\end{itemize}

\begin{figure*}[!htbp]
    \centering
	\includegraphics[width=.95\linewidth]{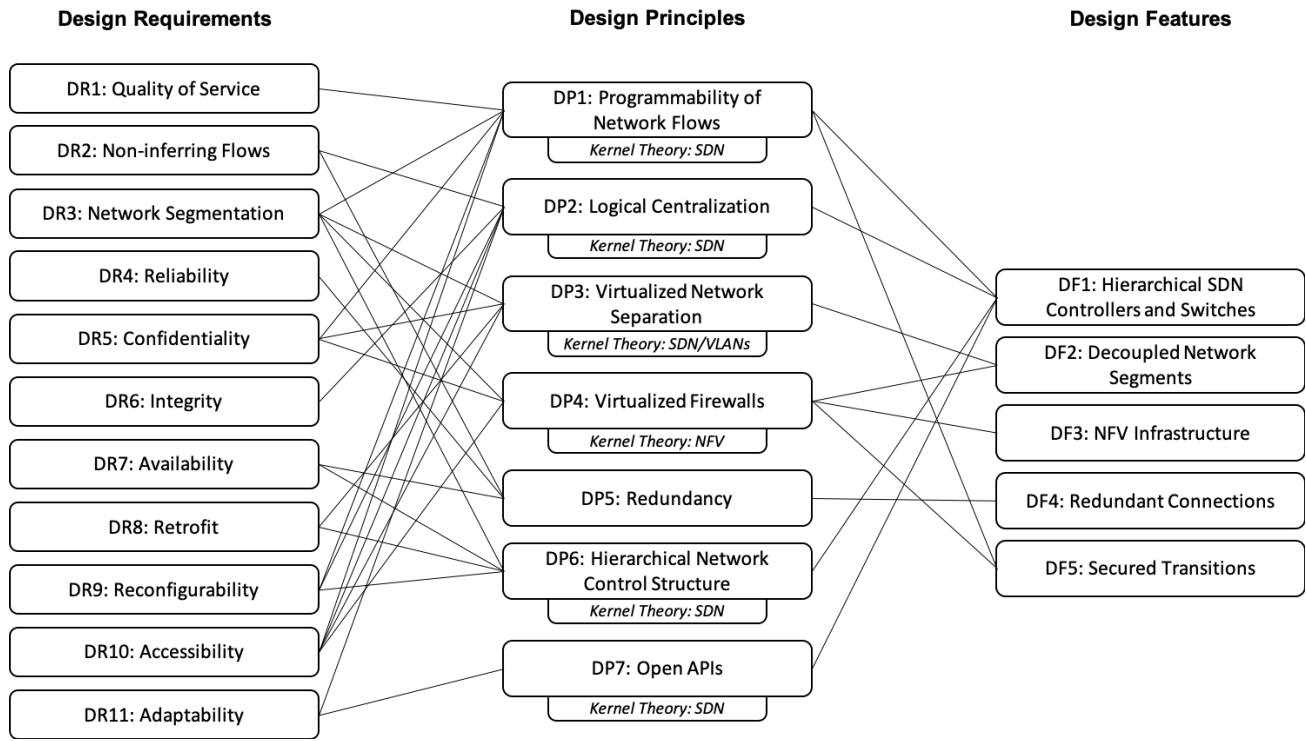}
	\caption{Mapping of design requirements derived from literature and practice to design principles and design features}
	\label{fig:mapping}
\end{figure*}

According to our overall goal of generating design knowledge for the design of a reference architecture for IIoT networks, we link conceptual aspects of the kernel theories SDN and NFV described in \Cref{sec:relatedwork} to the identified design requirements. \Cref{fig:mapping} summarizes this mapping visually and gives an overview of the selected design principles and their assignment to the design requirements. 

These principles include the programmability of network flows, logical centralization, network separation, a hierarchical network control structure and the support of open application programming interfaces (APIs) being central components of SDN as well as virtualized firewalls which can be realized by NFV and redundancy.

\begin{figure*}[!htbp]
    \centering
	\includegraphics[width=.95\linewidth]{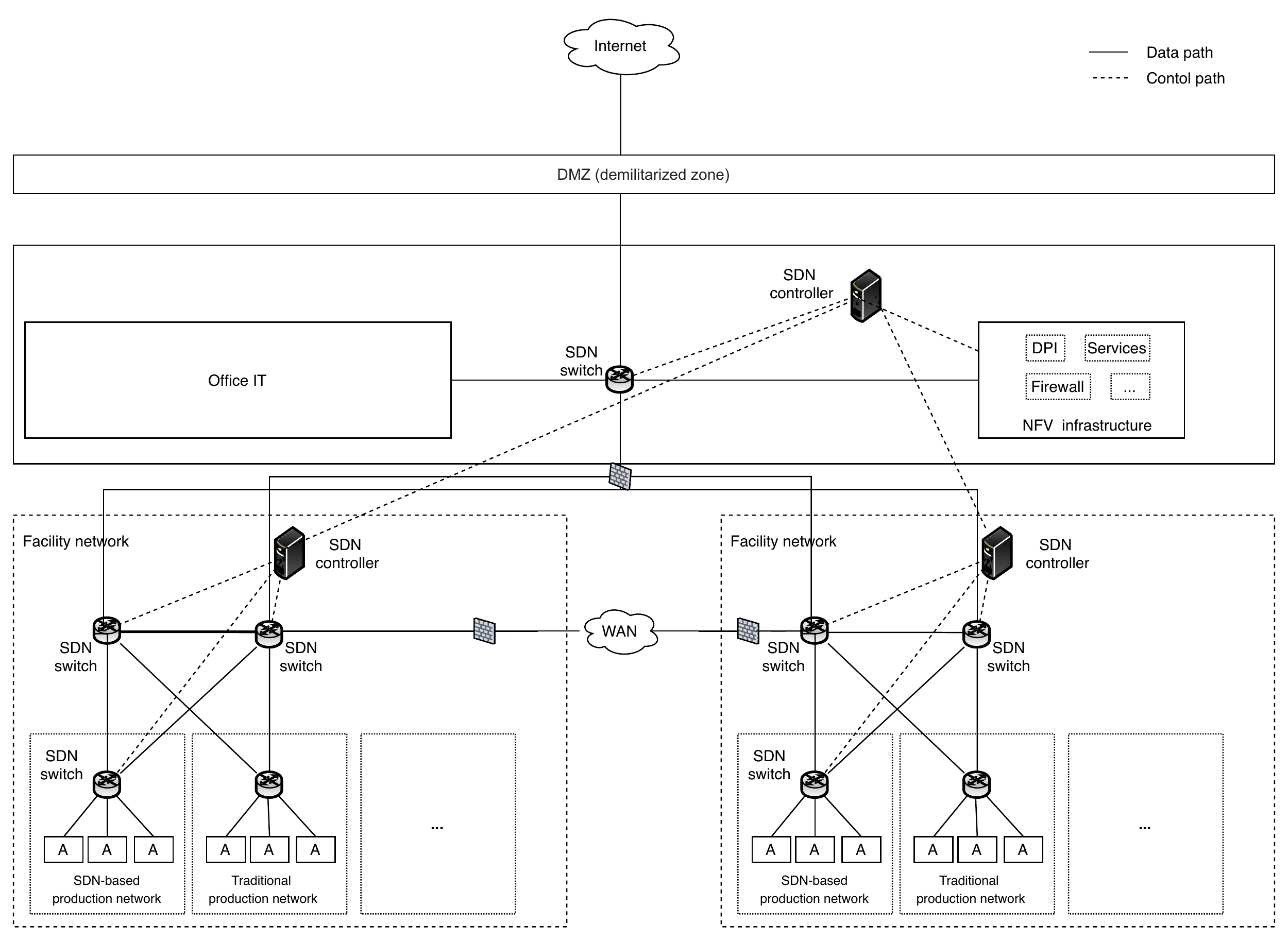}
	\caption{Reference architecture for future IIoT networks}
	\label{fig:architecture}
\end{figure*}

In a subsequent step, the design principles are translated into concrete design features which are incorporated into the implementation of the reference architecture. \Cref{fig:architecture} schematically depicts an exemplary implementation of a future oriented IIoT network and therefore serves as a blueprint for its actual realization in practice.

SDN switches provide the basic data forwarding functionalities in the network. The SDN switches are controlled by hierarchically assigned and individually programmable SDN controllers \emph{(DF1)}. Thus, these SDN controllers can either control the data flows of individual assets, or production lines (several assets are controlled by one SDN controller), or even the entire facility network \emph{(DF2)}. In areas where no SDN-capable hardware can be installed due to the age of the equipment or other circumstances, the existing technology is retained. The top-level SDN controller is responsible for orchestrating the NFV infrastructure throughout the company's IT infrastructure \emph{(DF3)}. Thus, NFV infrastructure services, such as segment-specific firewalls, can be provided to the entire network. Additionally, SDN switches are potentially connected through several transitions \emph{(DF4)}. Thus, data flows can either be steered according to individual network requirements or predefined rules based on different processes. Different QoS settings such as bandwidth, latency or jitter might be included in the routing decision of the SDN controller. Furthermore, different security policies are configured and enforced by the SDN controller allowing or blocking traffic within or across network segments \emph{(DF5)}.

\section{Evaluation}

\begin{figure}[!htbp]
    \centering
	\includegraphics[width=1\linewidth]{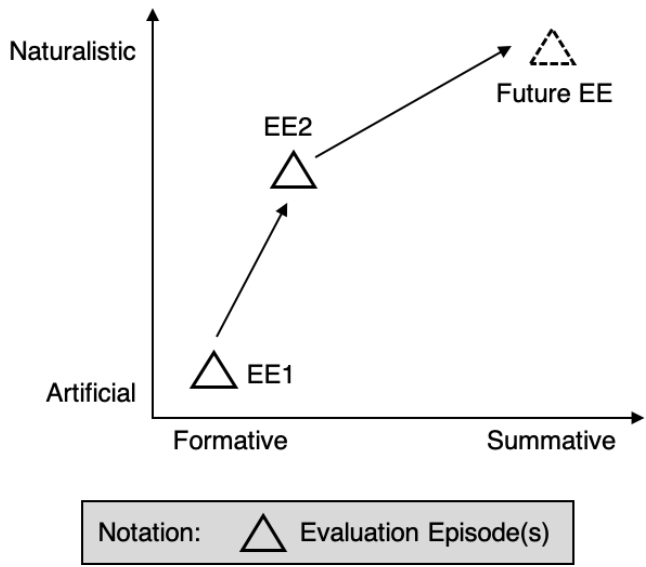}
	\caption{Overview of the evaluation strategy according to \citet{Venable2016} as well as the conducted and future evaluation episodes}
	\label{fig:feds}
\end{figure}

The artifact is evaluated in two separate episodes as depicted in \Cref{fig:feds}. According to \citet{Venable2016}, we perform the first episode in an artificial manner, leveraging logical arguments to verify the fulfillment of the initially formulated design requirements.

In general, the proposed SDN-based architecture supports different QoS settings \cite{Karakus2017}. However, especially in the area of time-critical communication, weaknesses may still be apparent, particularly when data streams have to be routed via the SDN controller. \citet{BenHadjSaid2019} describe an approach to resolve this issue by leveraging the time sensitive networking (TSN) protocol as an extension to SDN. Thus, design requirement \emph{DR1} is fulfilled.

Due to the SDN controller's total view and redundant interconnections, data flows can be dynamically rerouted through alternative routes, if an increased utilization of the transmission medium or components such as switches
or routers is detected \emph{(DR2)}.

Through the application of SDN and virtual local area networks (VLANs), the architecture enables assets and devices to be easily separated into different network areas. \citet{tsuchiya2018software} have shown that such a network separation can be supported by virtualized firewalls provided by a NFV infrastructure. Thus, network segmentation \emph{(DR3)} with secured transitions is enabled by the proposed reference architecture.

Reliability \emph{(DR4)} and availability \emph{(DR7)} is achieved through both redundant components (such as controllers, switches and transitions) and the hierarchical architecture. 
Based on NFV infrastructure, a user authentication of devices, as is already standard in most industrial networks today, can be implemented \emph{(DR5)}. \citet{yakasai2015flowidentity}, for instance, show how access control for the network can be implemented using SDN extended by the traditional 802.1X protocol.

While virtualized network functions offer great advantages in terms of flexibility, they also open up vulnerability to intrusion attacks. To counter this, \citet{Girtler2017} propose a mechanism that secures communication between the SDN controller and the network applications and thus enables integrity \emph{(DR6)}.

\citet{Amin2018} show that there are different ways to integrate existing IT infrastructure into an SDN-based network \emph{(DR8)}. However, these hybrid networks have the limitation that non-SDN-based network areas do not benefit from functionalities provided by SDN.

\citet{Kobzan2020, Madiwalar2019} demonstrate that IT devices or even conversion processes in the sense of `plug \& produce' are possible \emph{(DR9)}. Moreover, implementations have already been realized based on industrial use cases. However, an abstract information model for all properties of IT devices is required for a device's registration.

Due to the hierarchical SDN architecture combined with NFV, accessibility is ensured by default \emph{(DR10)}. NFV can be leveraged to enable access functions such as the provision of a virtual private network (VPN) or dedicated interfaces for external services. Moreover, SDN provides open interfaces and enables programmability of all network flows with the goal to keep the architecture as generic and adaptable as possible \emph{(DR11)}.

For a second evaluation episode we use exploratory focus groups \cite{tremblay2010focus} with the goal of discussing and evaluating the general feasibility of the proposed design features and the artifact itself \cite{pries2008strategies}. In total, 5 interviewees of a large enterprise in the automotive, industrial and building technology and IoT sectors participated in two sessions---one with an application focus and another one with a technical focus.

From the experts' point of view, the structure of the reference architecture and its main logical functional blocks are clearly presented and their functionalities cover the requirements identified. Only the description of the NFV infrastructure could have been more detailed to illustrate the multitude of realization possibilities. Moreover, in their opinion, the reference architecture as well as the underlying design requirements and features offer a great opportunity to discuss new areas of applications based on the use cases that could be realized leveraging them. Furthermore, doubts arose to some extent as to the practical suitability and maturity of the technologies included. Overall, all design features as well as the proposed architecture itself were perceived as feasible and useful for the implementation of future IIoT networks.

\section{Conclusion}

Building on the lack of future orientation of today's industrial IT networks of traditional manufacturing companies and hesitant innovation efforts due to outdated network infrastructures, this article aims at both systematically identifying requirements in literature and practice as well as deriving a reference architecture that can serve as a blueprint for future implementations.
Following the design science research paradigm, we approach this research effort in a structured manner by conducting a broad search for requirements in literature and practice. While related work only addresses isolated facets, we aim at a holistic picture and contribute a detailed and comprehensive list of requirements, derived design principles and features as well as a reference architecture covering the identified requirements. In addition, we pursue two separate evaluation episodes, which prove the general feasibility. 

For future research, a practical implementation remains as an additional evaluation episode. A further limitation is the selection of three use cases that represent different aspects of future industrial applications. However, these may not cover all potential applications. Furthermore, the reference architecture proposed is only one possible approach to support the design of future-proof IIoT networks. Due to the focus on SDN and the disregard of alternative solutions, it remains to be seen whether alternative approaches are conceivable as well. Thus, it is still a long way to go until rigid network architectures no longer slow down innovative data-based applications and business models. Beyond that, however, we are confident that this article will make a substantial contribution to the dissemination of modern network architectures and implementations. 

\section*{Acknowledgment}
This work has been supported by the German Federal Ministry of Education and Research through the research project ``bi.smart'' (grant no. 02J19B041).

\bibliographystyle{IEEEtranN}
\bibliography{references}

\onecolumn

\section*{Appendix}

\renewcommand{\thetable}{A\arabic{table}}
\setcounter{table}{0}


\begin{table}[h!]
\centering
\caption{Requirements derived from literature and practical use cases}
\label{tab:requirements}
\small


\begin{tabular}{lp{11.5cm}ll}
\textbf{No} & \textbf{Requirement}                                                                                                           & \textbf{Source} & \textbf{DR} \\ 
1           & The system should provide functions that ensure system integrity                                                               & Literature      & 6                           \\
2           & The system should provide functions that ensure the integrity of information                                                   & Literature      & 6                           \\
3           & The system should be able to detect security breaches                                                                          & Literature      & 5                           \\
4           & The system should be able to guarantee data confidentiality                                                                    & Literature      & 3,5                        \\
5           & The system should be able to restrict the data flow                                                                            & Literature      & 3                           \\
6           & The system should be able to support flexible data provision                                                                   & Literature      & 10                          \\
7           & The system should be able to compensate for failures of network components                                                     & Literature      & 4                           \\
8           & The system should be able to perform network segmentation                                                                      & Literature      & 3                           \\
9           & The system should support functions for event recording                                                                        & Literature      & 11                       \\
10          & The system should be able to implement role-based access controls                                                              & Literature      & 5                           \\
11          & The system should be able to implement role-based usage controls                                                               & Literature      & 5                           \\
12          & The system should support functions for perimeter protection                                                                   & Literature      & 10                          \\
13          & The system should support remote access (e.g., via VPN) functions                                                               & Literature      & 10                          \\
14          & The system should be able to implement different security policies for different network areas                                 & Literature      & 3,5                         \\
15          & The system should be able to implement network protection measures                                                             & Literature      & 3,4,5                       \\
16          & The system should provide high availability for operational technology (OT)                                                    & Literature      & 1,7                         \\
17          & The system should provide real-time communication for operational technology (OT) devices                                      & Literature      & 1                           \\
18          & The system should be able to scale and support as many devices as possible                                                     & Literature      & 1,4,9                       \\
19          & The system should be able to implement QoS (quality of service) guidelines                                                     & Literature      & 1                           \\
20          & The system should be able to increase network visibility                                                                       & Literature      & 10,11                          \\
21          & The system should be able to identify and block data flows                                                                     & Literature      & 3,5                         \\
22          & The system should offer the possibility of providing time-guaranteed communication with a defined pattern                      & Literature      & 1                           \\
23          & The system should support edge computing for local process processing                                                          & Literature      & 8,9                         \\
24          & The system should be able to support deterministic cyclic data communication                                                   & Literature      & 1                           \\
25          & The system should be able to separate time-critical from non-time-critical communication                                       & Literature      & 1,2                        \\
26          & The system should be able to support different communication standards                                                         & Literature      & 11                          \\
27          & The system should support functions that connect new field devices to the network without human intervention                   & Practice         & 8,9                         \\
28          & The system should be able to connect to an edge device                                                                         & Practice         & 9                           \\
29          & The system should be able to establish communication between an edge device and services outside the same zone                 & Practice         & 3,5,10                      \\
30          & The system should be able to dynamically provide different users (humans, machines, services) with direct access to the camera & Practice         & 1,5                         \\
31          & The system should be able to transmit camera data in real time                                                                 & Practice         & 1                           \\
32          & The system should be able to forward image data without hindering time-critical communication                                  & Practice         & 1,2                        \\
33          & The system should be able to allow a dynamic configuration of the camera                                                       & Practice         & 9                           \\
34          & The system should be able to provide sufficient data throughput for processing large amounts of data                           & Practice         & 1                           \\
35          & The system should be able to provide direct access to machine components                                                       & Practice        & 1,5                       \\
36          & The system should be able to provide role-based access to different network areas                                              & Practice        & 5                           \\
37          & The system should be able to maintain a communication link between the two end points                                          & Practice        & 1,7                         \\
38          & The system should be able to implement safety relevant guidelines                                                              & Practice        & 3                           \\
39          & The system should provide sufficient data throughput for processing large amounts of data                                      & Practice        & 1                           \\
40          & The system should be able to dynamically adapt the communication link                                                          & Practice        & 9                           \\
41          & The system should be able to support M2M communication                                                                         & Practice        & 9,11                        \\
42          & The system should provide functions that allow the location of devices in different network areas                              & Practice        & 3,9                         \\
43          & The system should be able to allow communication over different network transitions                                            & Practice        & 3                           \\
44          & The system should be able to establish an ad-hoc communication link between production machine and autonomous transport system & Practice        & 9                           \\ 
\end{tabular}


\end{table}

\end{document}